\documentstyle[12pt]{article}
\begin{document}
\renewcommand{\baselinestretch}{2}
\newcommand{\noi}{\noindent}
\newcommand{\be}{\begin{equation}}
\newcommand{\ee}{\end{equation}}
\newcommand{\bea}{\begin{eqnarray}}
\newcommand{\eea}{\end{eqnarray}}
\newcommand{\ba}{\begin{array}}
\newcommand{\ea}{\end{array}}
\newcommand{\half}{\frac{1}{2}}
\newcommand{\fourth}{\frac{1}{4}}
\newcommand{\kk}{k_{2}^{2}+k_{3}^{2}}
\newcommand{\km}{k_{1}^{2}+m^{2}}
\newcommand{\pic}{\scriptscriptstyle}
\newcommand{\esp}{\vspace{0.5cm}}

\newcommand{\nS}{\mbox{$n_{S}$}}
\newcommand{\nT}{\mbox{$n_{T}$}}
\newcommand{\xT}{\mbox{$x_{T}$}}
\newcommand{\CT}{\mbox{$C^{T}_{2}$}}
\newcommand{\CS}{\mbox{$C^{S}_{2}$}}
\newcommand{\Hpres}{\mbox{$100h\,{\rm km\,sec^{-1}\,Mpc^{-1}}$}}

\begin{center}
{\Large \bf WEYL EQUATION IN G\"ODEL TYPE UNIVERSES}
\end{center}

\vspace{.2in}
\begin{center}
Luis O. Pimentel\footnote{email: lopr@xanum.uam.mx}, Abel Camacho and Alfredo
Macias\footnote{email:amac@xanum.uam.mx } \\
{\it  Departamento de F\'{\i}sica Universidad Aut\'onoma
Metropolitana-Iztapalapa, Apartado Postal 55-534, M\'exico D. F., MEXICO.}

\vspace{.2in}
%CSR-AT-94-2 \\
%Submitted to {\it }
\end{center}

\begin{quote}
{\bf Abstract:}

The Weyl equation (massless Dirac equation) is studied in a family
of metrics of  the G\"odel type. The field equation
is solved exactly for one member of the family.

\end{quote}

\begin{center}
\vspace{.2in}
PACS numbers~~~04.20q,04.20.Jb,04.40+c,98.80.k\\
\end{center}

\vspace{.3in}

\def\square{\mathchoice\sqr54\sqr54\sqr{6.1}3\sqr{1.5}6\,\,}
\def\qa {^{\prime}}
\def\qq {^{\prime\prime}}
\def\qb {^{\prime 2}}
\def\noi {\noindent}
\baselineskip=20pt

\section {Introduction}

 Field equations for different spins have been studied in the
G\"odel universe~${}^{1-4}$%\cite{cvd,his,mash,pm}
 and its extensions~${}^{1,5-7}.$%\cite{cvd,ae,ds,vi}.
Here we consider the massless spinor
field equation in a family of homogeneous space-times of the
G\"odel type ${}^{8,9}$.%\cite{ray,chak}.
The metric of this class of space-
times is the cylindrically symmetric stationary type. We write
the metric in the following form,

\be
ds^2 =-(dt+m d\phi)^2 + dr^2+dz^2 +(l-m^2)d\phi^2,
\ee

\noi where $m$ and $l$ are functions of $r$, they also satisfy one
of the following conditions:

\be
D:= (l+m^2)^{1/2}= A_1 \exp [a r] +A_2 \exp [-a r], \;\;
\frac{1}{D}\frac{dm}{dr}=C
\ee

\noi or

\be
D=Ar ,\;\; \frac{1}{D}\frac{dm}{dr}=C,
\ee

\noindent here $ A_i$ and C are arbitrary constants. The whole matter
content in the universe described by metric (1) rotates uniformly and
intrinsically with local angular velocity:

\be
\Omega^\mu =\Bigl ( 0,0,0,\frac{-c}{\sqrt{-g}}\frac{dm}{dr}\Bigr )
\ee

\section {FIELD EQUATION}

In order to write the field equation for a massless spin 1/2 field we
introduce the tetrad $e^\alpha _{\ \mu}(x) $, that satisfies the usual
relation

\be
g_{\mu\nu}(x)=e^\alpha _{\ \mu}(x) e^\beta _{\ \nu}(x) \eta _{\alpha \beta}.
\ee

For the present case we can choose

\be
e^0_{\, 0}(x) =1,\quad e^1 _{\, 1}(x)= 1
, \quad  e^0 _{\,  2}(x) =\frac {-m}{D},\quad  e^2 _{\,  2}(x) =\frac
{1}{D},\quad e^3_{\, 3}(x) =1
\ee

\noi where the $x^0=t,x^1=r,x^2=\phi , x^3=z$.

The curved Dirac matrices that satisfy

\be
\gamma _\mu(x)  \gamma _\nu(x)+\gamma _\nu(x)\gamma _\mu(x) =-2
g_{\mu\nu}(x),
\ee

\noi are given by %[8-9]

\be
\gamma ^0(x) =\tilde\gamma ^0 -\frac {m}{D}\tilde \gamma^2,\quad \gamma
^1(x) =\tilde\gamma ^1,\quad
\gamma ^2(x)=\frac {1}{D}\tilde\gamma _2,\quad \gamma^3 (x)
=\tilde \gamma^3,
\ee

\noi where the $\tilde\gamma ^{\alpha}$ are Dirac matrices in flat Minkowski
spacetime.

Weyl equation in curved spacetime is

\be
\gamma ^{\mu} \nabla _\mu  \psi (x)=0,
\label{eq:dir}
\ee

\be
(1+\gamma ^{5} )\psi =0
\label{eq:chir}
\ee

\noi with

\be
\nabla _\mu =\partial _\mu + \Gamma _\mu,\quad \Gamma _\mu ={1\over 8}
[\tilde\gamma ^{\alpha}\tilde\gamma ^{\beta} ]e_\alpha ^{\,\nu} e_{\beta \nu
;\mu},
\ee

\noi and

\be
\gamma ^{5}=\gamma ^{0}\gamma ^{1}\gamma ^{2}\gamma ^{3}.
\ee

In the case of the metric (1) the spinorial connections are,

\bea
\Gamma_0=\frac{m'}{4D}\gamma^2 \gamma^1,\quad \Gamma_1=\frac{m'}{4D}\gamma^0
\gamma^2   ,\nonumber \\ \quad \Gamma_2=\frac{m'}{4}\gamma^0
\gamma^1 +\frac{mm'+l'}{4D}\gamma^1\gamma^3,\quad \Gamma_3=0
\eea

\noindent where the prime means derivative with respect to r. Using the
standard representation of the gamma matrices, Dirac equation reduces to

\be
\gamma^0 \psi_{,0} + \gamma^1 \psi_{,1} + {1\over D} \gamma^2 \psi_{,2}
+ \gamma^3 \psi_{,3} - {m\over D}\gamma^2 \psi_{,0}
-{m'\over 4 D}\gamma^2 \gamma^0 \gamma^1 \psi +\frac {2 D'}{D}
\gamma^1 \psi=0,
\ee

\noi in the above equation we can be separate variables according to

\be
\psi _{\bf k} =\exp{[-i(k_0 t+k_2 \phi +k_3 z)]}
\pmatrix{\xi _1(r)\cr
\xi_2(r)\cr}
\ee

\noi with $\xi _i$ two-component spinors, and $k_i$
constants. The chirality condition,
Eq.(\ref{eq:chir}) implies that

\be
\xi _1=\xi _2=\pmatrix{R _1 (r)\cr
R _2 (r)}
\ee

\noi and Eq.(\ref{eq:dir}) is now

\be
R_{2,1} + {\over D}(k_0 m-k_2)R_2 + i({m'\over 4 D}
-k_0-k_3)R_1=0,
\ee

\be
R_{1,1} - {1\over D}(k_o m - k_2)R_1 + i(k_3-k_0
-{m'\over 4 D})R_2=0.
\ee

\noindent In all the previous equations we have not used an
explicit form for the functions $l(r)$ and $m(r)$, we have been
able to solve Dirac equation for the second form of those
functions (Eq.(3)). If we eliminate $R_2$ between the above two equations
and using
the explicit expression for $ m=m_0+\frac{CA}{2} r^2$ we obtain

\be
R''_1+\bigl \{ \frac{1}{A r^2}(\tilde k_0 r^2- \tilde k_2)[1-
\frac{1}{A
}(\tilde k_0 r^2- \tilde
k_2)]+ E\bigr \} R_1 = 0,
\label{eq:fin}
\ee

\noi where we defined
\be
\tilde k_0={k_0 AC\over 2}, \tilde k_2=k_0 m_0-k_2, E=(k_3 - k_0-
{C\over 4})({C\over 4}-k_0 -k_3)-{2\tilde k_0 \over A}
\ee
\section { EXACT SOLUTIONS}

\noi The solution to Eq.(\ref{eq:fin}) is

\be
R_1 (r)=(\sqrt{2} \tilde A r^2)^{{-1\over 4}}[ A_1 M_{\kappa,\mu}(\sqrt{2}
\tilde A r^2) + A_2 W_{\kappa,\nu}(\sqrt{2} \tilde A r^2)],
\ee

\noi where

\be
 \tilde A= ({ k_0 \over A}),
\ee

\noi and
$M_{\kappa,\mu}$ and $W_{\kappa,\mu}$ are the Whittaker functions
defined in terms of the confluent hypergeometric functions  M and U,

\be
M_{\kappa,\mu}(x)= e^{-x/2} x^{1/2+\mu } M(\frac{1}{2}+\mu -\kappa,1+2\mu ,x),
\ee

\be
W_{\kappa,\mu}(x)= e^{-x/2} x^{1/2+\mu } U(\frac{1}{2}+\mu -\kappa,1+2\mu ,x),
\ee

\noi and the parameters $\kappa$ and $\mu$ in this case are given
by

\be
\kappa={E\over \sqrt{8} \tilde A},\;\;
\mu=\sqrt{{1\over 16} +{1\over 2}{1\over A} \tilde k_2 (1+{1\over A}
k_2)}
\ee

On this solution we impose the condition that the Whittaker
functions must be bounded for all values of $r$. This condition is
realized if and only if

\be
\frac{1}{2}+\mu -\kappa= -n%negative integer or zero.
\ee

noi here n is a positive integer or zer. From the above definitions that is
equivalent to

\be
\frac{1}{2}+\sqrt{{1\over 16} +{1\over 2}{1\over A} \tilde k_2
(1+{1\over A}
k_2)}-{E\over \sqrt{8} \tilde A}= -n ,
\ee

\noi where n is a natural number or zero.

In this way we have been able to solve the Weyl equations for a family of
spacetimes of the G\"odel type.

\section {ACKNOWLEDGMENTS}

This work was partially supported by the CONACYT GRANTS 1861-E9212,
3544-E9311.

\end{document}